\documentclass[10pt,aps,prl,twocolumn]{revtex4}
\usepackage{epsfig}

\begin{document}

\author{K.P. Santo and K.L. Sebastian \\
Department of Inorganic and Physical Chemistry\\
Indian Institute of Science\\
Bangalore 560012 \\
India.}
\title{A simple model for the kinetics of packaging of DNA in to a capsid against
an external force }

\begin{abstract}
We propose a simple model for the kinetics of packaging of viral DNA in to a
capsid against an external force trying to prevent it. The model leads to a
Butler-Volmer type dependence of the rate of packaging on the pulling force
F.
\end{abstract}

\maketitle

\begin{figure}[b]
\begin{center}
     \epsfig{file=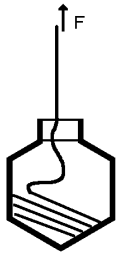, width=0.3\linewidth} 
     \caption{Packaging of the DNA in to the capsid against an external force $F$ trying to prevent it} 
\label{fig1}
\end{center}
\end{figure}

Recently, in a very interesting experiment\cite{Smith}, the rate of
packaging of a long viral DNA in to a capsid by means of the portal complex,
under the influence of an external force trying to prevent the packaging has
been studied (see Fig. \ref{fig1}). The force-velocity $(F-\nu )$ curves
show that packaging rate decreases even for small forces, implying that the
rate determining step in the packaging is affected by the externally applied
force. 
Using a Kramer's type model, involving steps that involve thermal activation
over a barrier, the experimental data was fitted assuming two force
dependent steps. The first involves a conformational change of only 0.1 nm
and is the rate determining step for small forces. At higher forces, the
velocity is found to decrease more sharply and it was suggested that the
second step, which has a larger conformational change associated with it,
becomes the rate-limiting step. In the following, we present a simple
theoretical model for the $F-\nu $ curves, which seems to fit the data
reasonably well. In our model, there is only one step in the activation
process, and we include the possibility of the reverse process too.
Interestingly the model leads to a formula analogous to the Butler-Volmer
equation of electrochemistry. A mechanism consistent with the structure of
the packaging motor, determined from X-ray crystallography has been
suggested \cite{Simpson}. According to this, the packaging involves
successive firing of five ATPases in the portal complex. The hydrolysis of
one ATP molecule leads to a movement of two base pairs of the DNA in to the
capsid. To analyze the process, using the methods of physical kinetics, we
need to have an idea of of the potential energy surface for the processes,
and of the transition state. Here, it is most convenient to take the
reaction co-ordinate to be the length of the DNA molecule inside the capsid,
which we shall denote as $x$. Every increase in $x$ by $a(=6.8\AA )$
involves the hydrolysis of one ATP molecule. This hydrolysis is coupled
mechanically to the movement of the DNA, through the ATPase. Therefore, as $%
x $ changes by $a$, there is a lowering of net free energy of the system. 
\begin{figure}[h]
\epsfig{file=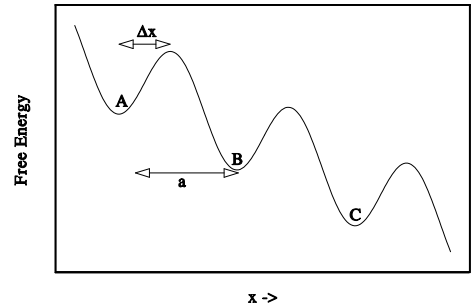, width=0.95\linewidth}
\caption{Free energy profile for the packaging. x denotes the length
of the DNA that has been packaged. The packaging of a length $a$
corresponds to going from A to B (or B to C) and involves the hydrolysis of
one ATP molecule. As more and more DNA gets packed in the oscillations in
the free energy profile would decrease and the profile would get flatter
and flatter. }
\label{fig2}
\end{figure} 

A free energy profile for the process, against the reaction co-ordinate ($x$%
) is shown in the Fig \ref{fig2}. Each minimum (A,B,C) in the curve is an
initial state for the packaging of two more base pairs, and has one ATP
molecule, ready to be hydrolyzed. In going from A to B, the ATP is
hydrolyzed to ADP (leading to a lowering of free energy) and two base pairs
have been packed in. At B, the ADP is got rid off, and another ATP is
attached (not to the same ATPase, see ref. \cite{Simpson}) and then the
system goes over from B to C. The experiments are done under saturating
concentration of ATP. Hence ATP attachment is not the rate determining step. 
\begin{figure}[h]
\epsfig{file=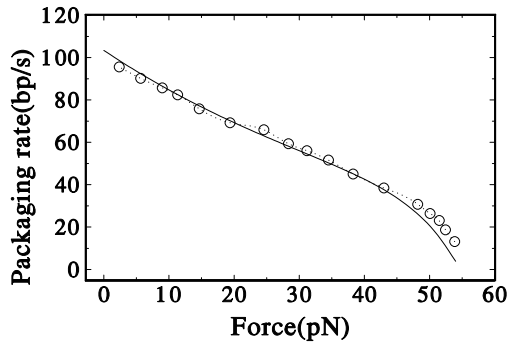,width=\linewidth}
\caption{Packaging rate $\nu$ against the force.
 The best fit was obtained using the non-linear fit package of MATHEMATICA 
and gave
 $ \nu_{0f}=103.41 bp/s $,
 $\nu_{0b}=0.0129bp/s $ and $\Delta x=0.821\AA $.   The data were taken from the figure of 3b of reference \cite{Smith} and corresponds to a packing of $1/3^{rd}$ of the capsid}
\label{fig3}
\end{figure} 
The transition state is at a distance of $\Delta x$ from A and therefore, in
presence of an external force $F$, to reach it one has to do an extra work
of $F\Delta x$. Therefore the rate of the forward process, in presence of
the force is given by $\nu _{0f}\exp (-F\Delta x/kT)$, where $\nu _{0f}$ is
the rate in the absence of the external force. The rate of the backward
reaction is, in a similar fashion, given by $\nu _{0b}\exp (F(a-\Delta x)/kT)
$, where $a=6.8\AA $. Hence the net rate is $\nu =\nu _{0f}\exp (-Fa\alpha
/kT)-\nu _{0b}\exp (Fa(1-\alpha )/kT)$, where $\alpha =\Delta x/a$. Using
this expression, we have fitted the $F-\nu $ curve in ref. \cite{Smith}, and
the result is shown in Fig. \ref{fig3}. It seems to fit the data fairly
well. Ref. \cite{Smith}. assumes two different rate determining steps,
operating at different force strengths. Further, they have four unknown
parameters which are used for fitting. In comparison, in our mechanism,
there is only one (reversible step) and the sudden decrease in the velocity,
observed around 45 pN is the result of the rate of reverse step becoming
comparable to that of the forward process. It has only three unknown
parameters. Interestingly, the above equation is very similar to the
Butler-Volmer equation of electrochemistry\cite{Atkins}, in which the force $%
F$ plays a role similar to the overpotential, and $\alpha $ is the analogue
of the symmetry factor. The equation can be used to calculate the force at which the packaging velocity would become zero and we find it to be 55.4 pN.

K.P. Santo is thankful to CSIR (India), for financial support.

\end{document}